\documentclass[12pt]{article}

\usepackage{epsf}
\usepackage[dvips]{graphicx}
\usepackage{amssymb}
\usepackage{cite}
\begin{document}

\begin{center}
{\bfseries CROSS SECTION IN DEUTERON-PROTON ELASTIC SCATTERING AT 1.25 GeV/u}

\vskip 5mm

P.K. Kurilkin$^{6 \dag}$, G.~Agakishiev$^{6}$, A.~Balanda$^{3}$, D.~Belver$^{15}$, A.~Belyaev$^{6}$, A.~Blanco$^{2}$, 
M.~B\"{o}hmer$^{11}$, J.~L.~Boyard$^{13}$, P.~Cabanelas$^{15}$, E.~Castro$^{15}$, S.~Chernenko$^{6}$, 
J.~D\'{\i}az$^{15}$, A.~Dybczak$^{3}$, E.~Epple$^{11}$, L.~Fabbietti$^{11}$, O.~Fateev$^{6}$, 
P.~Finocchiaro$^{1}$, P.~Fonte$^{2,17}$, J.~Friese$^{11}$, I.~Fr\"{o}hlich$^{7}$, T.~Galatyuk$^{7}$, 
J.~A.~Garz\'{o}n$^{15}$, A.~Gil$^{16}$, M.~Golubeva$^{10}$, D.~Gonz\'{a}lez-D\'{\i}az$^{4}$, 
F.~Guber$^{10}$, T.~Hennino$^{13}$, R.~Holzmann$^{4}$, P.~Huck$^{11}$, A.~Ierusalimov$^{6}$, 
I.~Iori$^{9,19}$, A.~Ivashkin$^{10}$, M.~Jurkovic$^{11}$, B.~K\"{a}mpfer$^{5,18}$,
T.~Karavicheva$^{10}$, I.~Koenig$^{4}$, W.~Koenig$^{4}$, B.~W.~Kolb$^{4}$, A.~Kopp$^{8}$, G.~Korcyl$^{3}$,
GK~Kornakov$^{15}$, R.~Kotte$^{5}$, A.~Kozuch$^{3,20}$, A.~Kr\'{a}sa$^{14}$, F.~Krizek$^{14}$,
R.~Kr\"{u}cken$^{11}$, H~Kuc$^{3}$, W.~K\"{u}hn$^{8}$, A.~Kugler$^{14}$, A.~Kurepin$^{10}$,
A.~Kurilkin$^{6}$, P.~KŠhlitz$^{5}$, V.~Ladygin$^{6}$, J.~Lamas-Valverde$^{15}$, 
S.~Lang$^{4}$, K.~Lapidus$^{10}$, T.~Liu$^{11}$, L.~Lopes$^{2}$, M.~Lorenz$^{7}$, 
L.~Maier$^{11}$, A.~Mangiarotti$^{2}$, J.~Markert$^{7}$, V.~Metag$^{8}$, B.~Michalska$^{3}$,
J.~Michel$^{7}$, C.~M\"{u}ntz$^{7}$, L.~Naumann$^{5}$, Y.~C.~Pachmayer$^{7}$, M.~Palka$^{7}$, 
Y.~Parpottas$^{12}$, V.~Pechenov$^{4}$, O.~Pechenova$^{7}$, J.~Pietraszko$^{7}$, W.~Przygoda$^{3}$,
B.~Ramstein$^{13}$, A.~Reshetin$^{10}$, J.~Roskoss$^{8}$, A.~Rustamov$^{4}$, A.~Sadovsky$^{10}$,
P.~Salabura$^{3}$, A.~Schmah$^{11}$, J.~Siebenson$^{11}$, Yu.~G.~Sobolev$^{14}$, S.~Spataro$^{8,21}$, 
H.~Str\"{o}bele$^{7}$, J.~Stroth$^{7,4}$, C.~Sturm$^{4}$, M.~Sudol$^{13}$, A.~Tarantola$^{7}$,
K.~Teilab$^{7}$, P.~Tlusty$^{14}$, M.~Traxler$^{4}$, R.~Trebacz$^{3}$, H.~Tsertos$^{12}$, T.~Vasiliev$^{6}$,
V.~Wagner$^{14}$, M.~Weber$^{11}$, J.~W\"{u}stenfeld$^{5}$, S.~Yurevich$^{4}$ and Y.~Zanevsky$^{6}$

\vskip 5mm

{\small
(1) {\it
Istituto Nazionale di Fisica Nucleare - Laboratori Nazionali del Sud, 95125~Catania, Italy
}
\\
(2) {\it
LIP-Laborat\'{o}rio de Instrumenta\c{c}\~{a}o e F\'{\i}sica Experimental de Part\'{\i}culas, 3004-516~Coimbra, Portugal
}
\\
(3) {\it
Smoluchowski Institute of Physics, Jagiellonian University of Cracow, 30-059~Krak\'{o}w, Poland
}
\\
(4) {\it
GSI Helmholtzzentrum f\"{u}r Schwerionenforschung GmbH, 64291~Darmstadt, Germany
}
\\
(5) {\it
Institut f\"{u}r Strahlenphysik, Forschungszentrum Dresden-Rossendorf, 01314~Dresden, Germany
}
\\
(6) {\it
Joint Institute of Nuclear Research, 141980~Dubna, Russia
}
\\
(7) {\it
Institut f\"{u}r Kernphysik, Goethe-Universit\"{a}t, 60438 ~Frankfurt, Germany
}
\\
(8) {\it
II.Physikalisches Institut, Justus Liebig Universit\"{a}t Giessen, 35392~Giessen, Germany
}
\\
(9) {\it
Istituto Nazionale di Fisica Nucleare, Sezione di Milano, 20133~Milano, Italy
}
\\
(10) {\it
Institute for Nuclear Research, Russian Academy of Science, 117312~Moscow, Russia
}
\\
(11) {\it
Physik Department E12, Technische Universit\"{a}t M\"{u}nchen, 85748~M\"{u}nchen, Germany
}
\\
(12) {\it
Department of Physics, University of Cyprus, 1678~Nicosia, Cyprus
}
\\
(13) {\it
Institut de Physique Nucl\'{eaire (UMR 8608), CNRS/IN2P3 - Universit\'{e} Paris Sud, F-91406~Orsay Cedex, France
}
\\
(14) {\it
Nuclear Physics Institute, Academy of Sciences of Czech Republic, 25068~Rez, Czech Republic
}
\\
(15) {\it
Departamento de F\'{\i}sica de Part\'{\i}culas, Univ. de Santiago de Compostela, 15706~Santiago de Compostela, Spain
}
\\
(16) {\it
Instituto de F\'{\i}sica Corpuscular, Universidad de Valencia-CSIC, 46971~Valencia, Spain
}
\\
(17) {\it
Also at ISEC Coimbra, ~Coimbra, Portugal
}
\\
(18) {\it
Also at Technische Universit\"{a}t Dresden, 01062~Dresden, Germany
}
\\
(19) {\it
Also at Dipartimento di Fisica, Universit\`{a} di Milano, 20133~Milano, Italy
}
\\
(20) {\it
Also at Panstwowa Wyzsza Szkola Zawodowa, 33-300~Nowy Sacz, Poland
}
\\
(21) {\it
Also at Dipartimento di Fisica Generale, Universita' di Torino, 10125 ~ Torino, Italy
}
\\
$\dag$ {\it
E-mail: pkurilkin@jinr.ru
}
}}
\end{center}

\vskip 5mm

\begin{center}
\begin{minipage}{150mm}
\centerline{\bf Abstract}
First results of the differential cross section in $dp$ elastic scattering at 1.25 GeV/u measured with the HADES over a large angular range are reported. The obtained data correspond to large transverse momenta, where a high sensitivity to the two-nucleon and three-nucleon short-range correlations is expected.
\end{minipage}
\end{center}

\vskip 10mm

\section{Introduction}
\hspace*{\parindent}The investigation of nuclear reactions involving deuterons has played an important role in the development of our understanding of the nuclear structure and the dynamics of nuclear interactions. Nowadays, the investigation of processes with the participation of a deuteron is undoubtedly useful to solve problems, in particular relativistic ones.\\
\hspace*{\parindent}To date, a large volume of data on the deuteron has been accumulated from the experiments with electron and hadron beams. A feature of the modern investigations of the deuteron by electron beams~\cite{ref_1,ref_2,ref_3} is that they are focused on revealing the deuteron internal structure at relatively small distances. The study of hadron reactions, in particular the collisions of relativistic deuterons with nucleons and nuclei also provide important information on the deuteron structure at small distances. Moreover, the study of collisions of relativistic deuterons with nucleons and nuclei is extremely important for clarifying fundamental problems of the relativistic description of fast-moving composite objects.\\
\hspace*{\parindent}The deuteron is the only bound state of two-nucleon system. The deuteron wave function $\Psi^{JM}({\bf r})$ can be expressed as
\begin{equation}
\Psi^{JM}({\bf r})= \frac{u(r)}{r}Y^{JM}_{01}(r) + \frac{w(r)}{r}Y^{JM}_{21}(r),
\end{equation} 
where u(r) and w(r) are the radial wave functions of the S and D states, and $Y^{JM}_{LS}$ are the spherical harmonics. The parameters of the deuteron wave function are determined by solving the Schr$\ddot o$dinger equation for the potentials constructed such as to describe the data on the nucleon-nucleon scattering at energies below the pion production threshold. Modern models of nucleon-nucleon potentias (e.g. Nijmegen, CD-Bonn, AV-18 etc.) reproduce the $NN$ scattering data up to 350 MeV with very good accuracy, including the deuteron static properties. At large distances between nucleons in a deuteron, these models yield almost identical results; they start to differ in their structure at small distances, predicting different high-momentum components of the deuteron wave function. 
The triton binding energy, unpolarized deuteron-proton elastic scattering and breakup data \cite{ref_7_src_ladygin} can be reproduced with incorporation of the three-nucleon forces (3NF)  incorporation only. However, the use of different 3NF models in Faddeev calculations can not reproduce polarization data intensively accumulated during the last decade at different facilities~\cite{ref_8_src_ladygin,ref_10_src_ladygin,ref_11_src_ladygin,ref_14_src_ladygin}.\\
\hspace*{\parindent}On the other hand, cross section data in $pd$ elastic scattering obtained already at the energy of 250 MeV~\cite{ref_11_src_ladygin} cannot be reproduced by Faddeev calculations with the modern 3NF inclusion. The authors~\cite{ref_11_src_ladygin} stated that the reason of this discrepancy may be the neglect of a new type of short-range 3NF.  \\
\hspace*{\parindent}Fundamental degrees of freedom in the framework of QCD, i.e. quarks and gluons, begin to play a role at internucleonic distances comparable with the size of the nucleon. At high energy and large transverse momenta the constituent counting rules (CCR)~\cite{ref_const_count_roles1,ref_const_count_roles2} yield a good results. They predict that, for a binary reaction, the differential cross section for a given scattering angle can be parametrized as
\begin{equation}
\frac{d\sigma}{dt}(AB\to CD)=\frac{f(t/s)}{s^{n-2}},
\end{equation}
where $n = N_{A}+N_{B}+N_{C}+N_{D}$ is the total number of fundamental constituents (quarks) involved in the reaction. Amplitude $f$ is a function of the scattering angle. The regime corresponding to CCR in the collision of light nuclei can occur already at $T_d=500$ MeV\cite{ref_uzikov}.\\
\hspace*{\parindent}Thus, the investigation of the energy dependence of the cross section in $dp$ elastic scattering is very desirable to obtain further information on the deuterons internal structure. 

\begin{figure}[t]
 \epsfysize=0.55\textwidth
 \centerline{
 \epsfbox{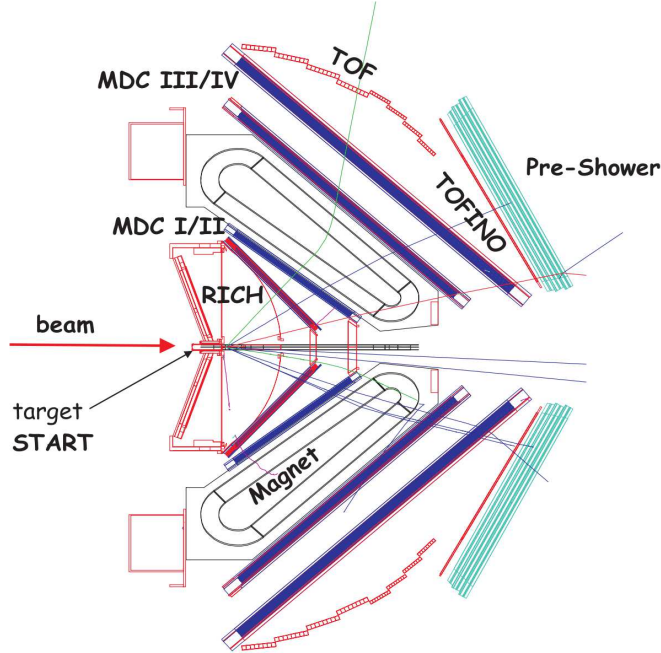}}
 \caption{Schematic view of the HADES detector. A RICH detector with gaseous radiator, carbon fiber mirror and UV photon detector with solid CsI photocathode is used for electron identification. Two sets of Mini-Drift Chambers (MDCs) with 4 modules per sector are placed in front and behind the toroidal magnetic field to measure particle momenta. A time of flight wall (TOF/TOFINO) combined with a Pre-Shower detector at forward angles is used for additional electron identification and trigger purposes. The target is placed at half radius off the centre of the mirror. For reaction time measurements, a START detector is located in front of the target. A few particle tracks are presented also.}
\label{fig1_hades}
\end{figure}

\section{HADES experiment}
\hspace*{\parindent}The High-Acceptance DiElectron Spectrometer (HADES) is in operation at GSI (Darmstadt). It has been specifically designed to study medium modification of the light vector mesons $\rho$, $\omega$, $\phi$~\cite{ref_2_geydar}. The HADES detector (Fig.~\ref{fig1_hades}) consists of six identical sectors covering almost full azimutal range and a polar angle ranges from $18^\circ$ up to $85^\circ$. Momentum measurement derives from the reconstruction of the particle trajectory using the four Mini-Drift Chambers (MDC) located before and after the magnetic field region. Electron identification is performed with the hadron-blind gas Ring Imaging Cherenkov detector (RICH) together with Time-of-Flight (TOF/TOFINO) and an electromagnetic pre-shower detectors (Pre-Shower). 
A powerful two-stage trigger system is employed to select events within a predefined charged particle multiplicity interval (first-level trigger LVL1), as well as electron candidates (second-level trigger LVL2). A detailed description of the main spectrometer components can be found in~\cite{ref_hades_detector_overview}.\\
\hspace*{\parindent}A dedicated physics program including heavy ions, deuteron, proton and pion beams, proposed for the HADES detector in~\cite{ref_4_geydar, ref_5_geydar}, is carried out.

\begin{figure}[t]
\begin{minipage}[t]{0.45\textwidth}
\centering
 \resizebox{0.95\textwidth}{!}{\includegraphics[width=0.95\textwidth,height=0.75\textwidth]{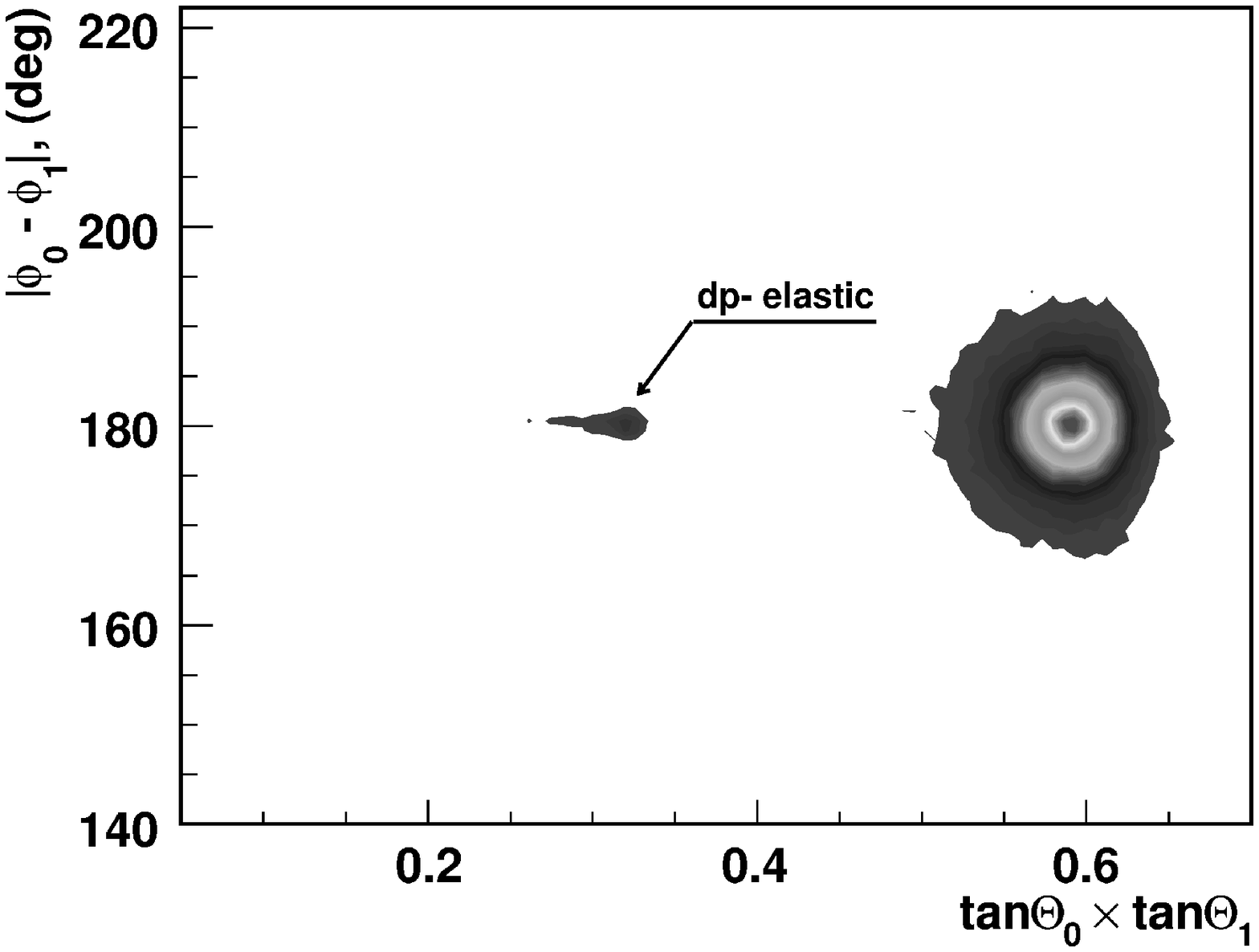}}
\caption{Correlation between the difference of azimuthal angles of tracks and their polar angles product for two charged particles. Left weak spot corresponds to the wanted $dp$ elastic scattering events.}
\label{fig_2_hades}
\end{minipage}
\hspace*{1.2cm}
\begin{minipage}[htbp]{0.45\textwidth}
\vspace*{-2.4cm}
\centering
 \resizebox{0.95\textwidth}{!}{\includegraphics[width=0.95\textwidth,height=0.75\textwidth]{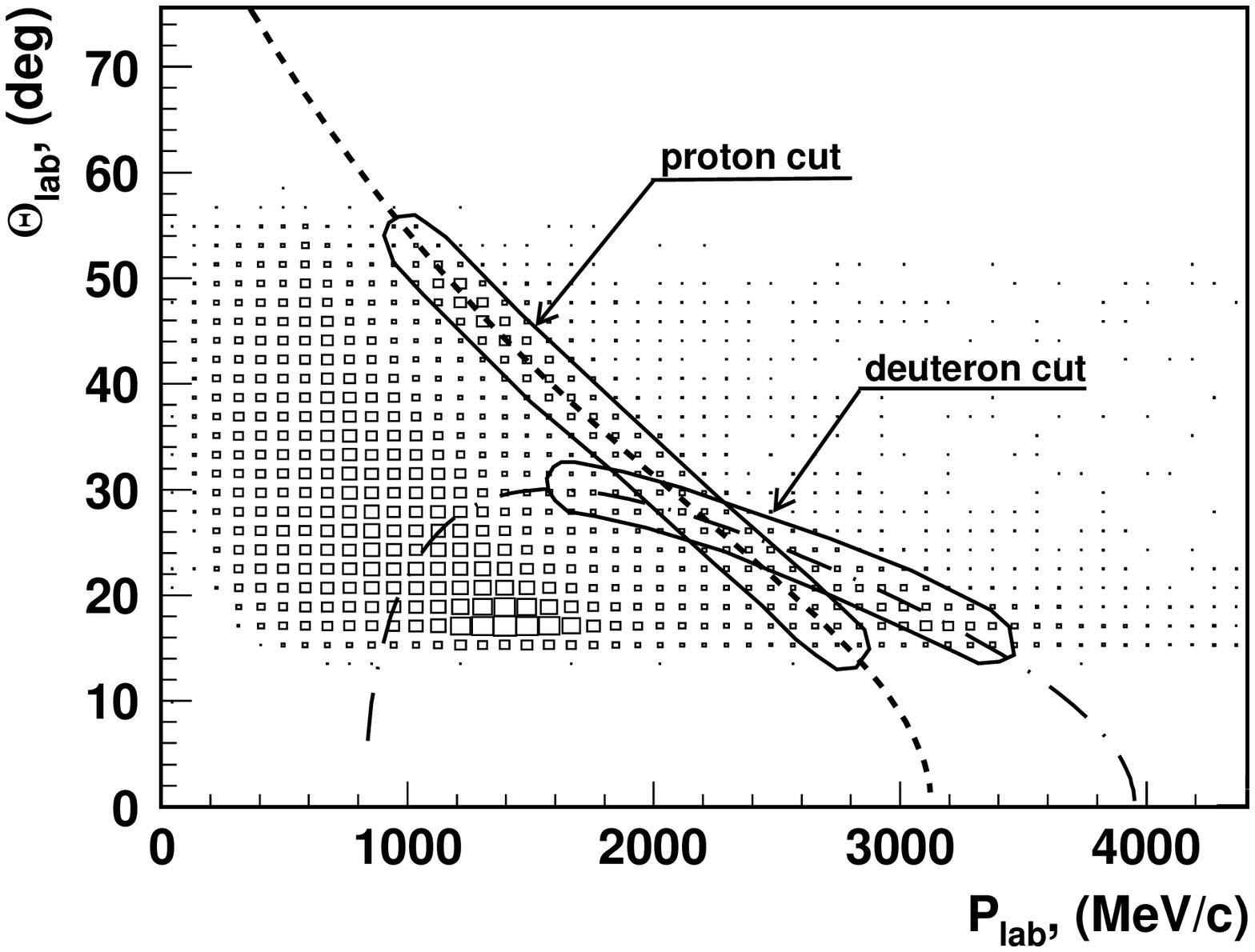}}
\caption{Angle-momentum correlation for $dp$ elastic scattering at 1.25 GeV/u. The graphical criteria for selecting deuterons and protons are shown by solid lines described in the text.}
\label{fig3_angle_mom_corr}
\end{minipage}
\end{figure}
In 2007, the HADES collaboration has performed a study of the $NN$ reactions in an experiment using deuteron-proton collisions. The main goal of this experimental run was to measure dielectron production properties in the $np$ channel at a kinetic beam energy of 1.25 GeV/u. In order to be capable separating the quasi-free $np$ channel, the HADES setup was upgraded with a Forward Wall (FW) scintillator hodoscope located 7 m downstream from  the target.

\begin{figure}[t]
\begin{minipage}[t]{0.45\textwidth}
\centering
 \resizebox{0.85\textwidth}{!}{\includegraphics[width=0.65\textwidth,height=0.57\textwidth]{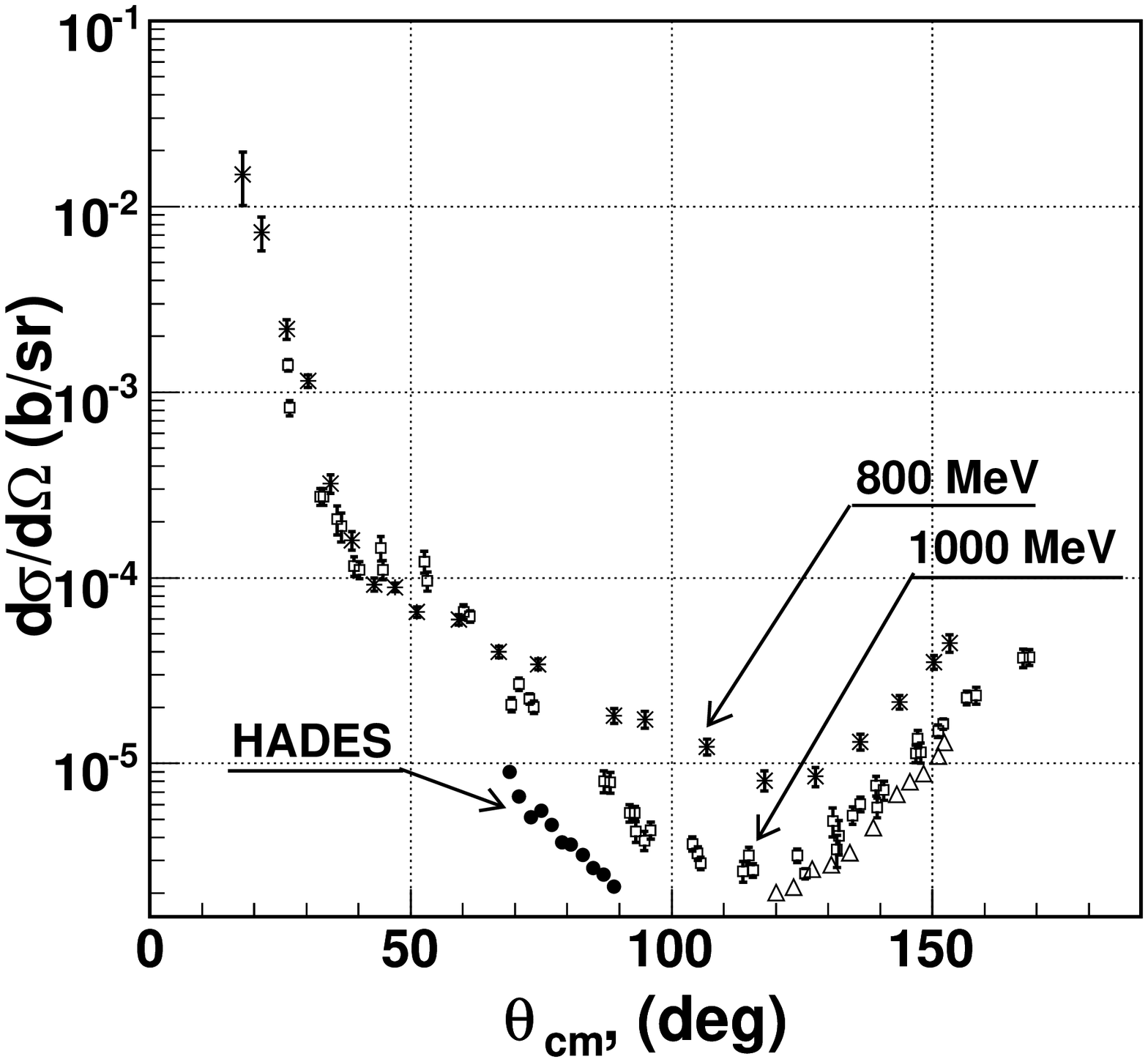}}
\caption{Differential cross section of $dp$ elastic scattering at 1.25 GeV/u. The world data obtained at 800 MeV and 1000 MeV are marked by $\ast$~\protect\cite{ref_1_data}, $\Box$~\protect\cite{ref_2_data} and $\triangle$~\protect\cite{ref_3_data},respectively. The preliminary HADES results are presented by full circles.}
\label{fig4}
\end{minipage}
\hspace*{.5cm}
\begin{minipage}[htbp]{0.45\textwidth}
\vspace*{-1.4cm}
\centering
 \resizebox{0.85\textwidth}{!}{\includegraphics[width=0.65\textwidth,height=0.57\textwidth]{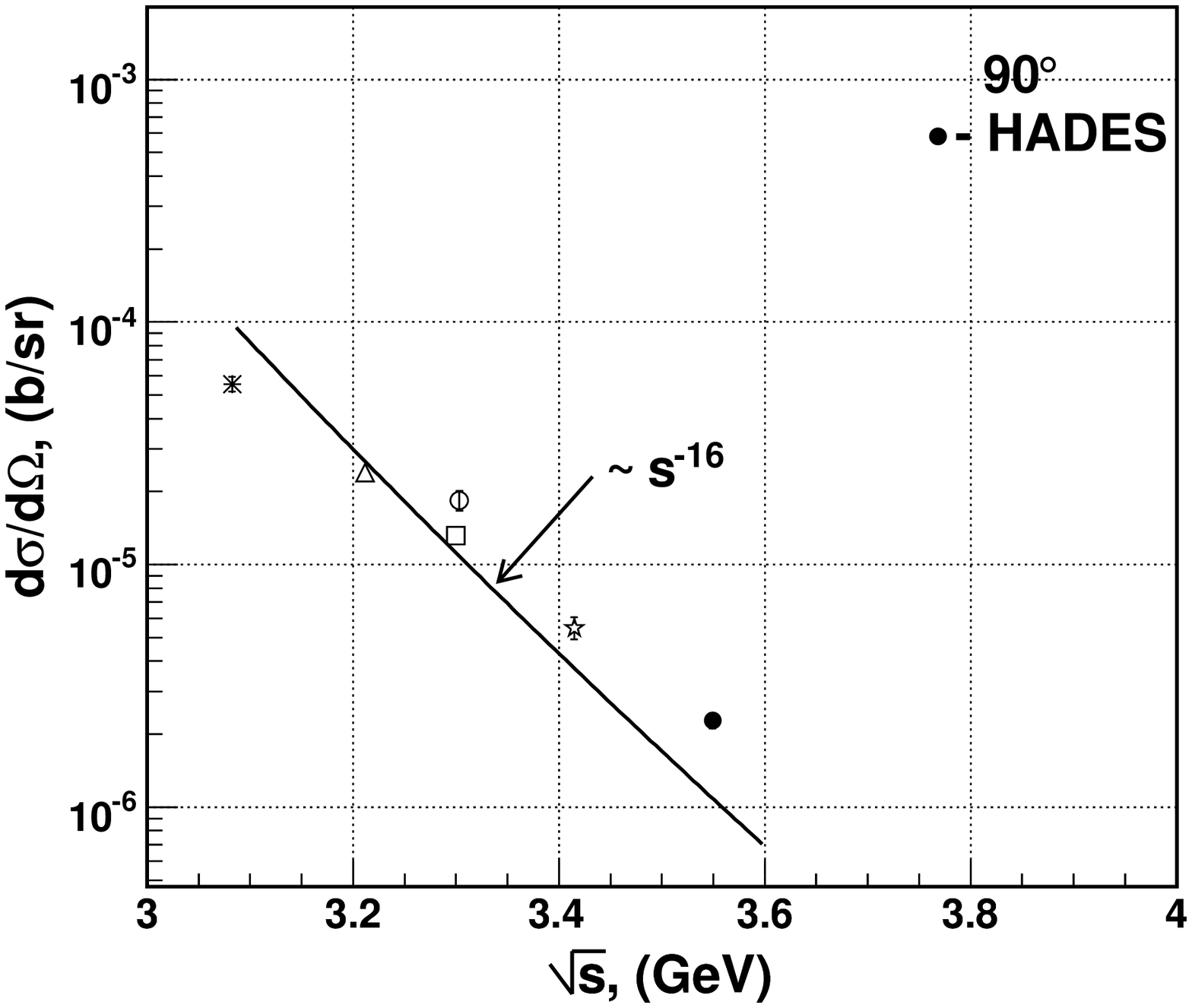}}
\caption{Differential cross section of $dp$ elastic scattering at the scattering angle of $90^\circ$ in c.m. The preliminary HADES results are presented by the full circle. The world data are marked by $\ast$~\protect\cite{ref_4_data}, $\triangle$~\protect\cite{ref_5_data},  $\Box$~\protect\cite{ref_5_data}, $\circ$~\protect\cite{ref_1_data} and $\star$~\protect\cite{ref_3_data}. The curve are the predictions of constituent counting rules.}
\label{fig5}
\end{minipage}
\end{figure}
The large acceptance of the HADES setup makes it also a unique tool to study the short-range correlations in channels without meson production. The study of the $dp$ elastic scattering at large scattering angles in the center of mass system can provide a new information about nucleon-nucleon interaction at small internucleonic distances.\\

\section{Results}
\hspace*{\parindent}In the experimental run a deuteron beam with 1.25 GeV/u kinetic energy was incident on a liquid hydrogen cell with a length of 5 cm. The intensity of the beam was $10^7$ particles/s.\\
\hspace*{\parindent}There are two conditions which can be used to select the $dp$ elastic pairs. The first one is related to the azimuthal angles $\phi_{0,1}$ denoting the difference of the angles of two particles labelled by 0 and 1. Coplanarity of the elastic proton and deuteron is expressed by $\mid\phi_{0}-\phi_{1}\mid = 180^\circ$. The second condition is related to the polar angles $\Theta$ of the particles.     
Fig.~\ref{fig_2_hades} represents the correlation between the difference of azimuthal angles of tracks and their polar angles product $tan\Theta_{0}\times tan\Theta_{1}$ for two charged particles labelled by 0 and 1. A weak spot in the left part of the figure corresponds to the $dp$ elastic scattering. The final selection of the deuteron-proton elastic events was made with the use of graphical criteria on the angle-momentum correlation of two tracks. Fig.~\ref{fig3_angle_mom_corr} shows the distributions of the polar angle of one elastically scattered particle versus its momentum compared to a  theoretical prediction. Dashed and dash-dotted curves correspond to the kinematical relationship between the polar angle and momentum for protons and deuterons in the laboratory frame, respectively. The polar angles and momenta of deuterons and protons from $dp$ elastic scattering are particularly the same in the region where graphical criteria overlap. Thus  the application of the graphical criteria on the angle-momentum correlation for $dp$ elastic scattering is not enough to separate the particles. The information about experimental time-of-flight difference for particles was used for the particle identification in the region where the graphical criteria overlap.\\
\hspace*{\parindent}Fig.~\ref{fig4} shows the differential cross section data in $dp$ elastic scattering versus scattering angle in c.m. in the GeV energy region. The existing world data are shown by the stars~\cite{ref_1_data}, boxes~\cite{ref_2_data} and triangles~\cite{ref_3_data}. The preliminary HADES results are represented by the full circles. These data cover a new kinematical region corresponding to high transverse momenta. On can see the tendency of the differential cross section to decrease with increasing the beam energy. 
The energy dependence of the $dp$ elastic scattering cross section data at the angle of $90^\circ$ is shown in Fig.~\ref{fig5}. The solid curve presented here was obtained by a world data fit with the function $f\sim 1/s^{16}$. The behavior of the cross section data at an energy of 1.25 GeV/u and at fixed scattering angle of $90^\circ$ in c.m. is in satisfactory agreement with the behaviour of the world data obtained at lower energies~\cite{ref_1_data,ref_3_data,ref_4_data,ref_5_data} as well as the constituent counting rules~\cite{ref_const_count_roles1,ref_const_count_roles2}.\\

\section{Conclusion}
\hspace*{\parindent}Preliminary results on the differential cross section in $dp$ elastic scattering are obtained at an energy of 1.25 GeV/u and large scattering angles in c.m.s. with the HADES spectrometer. A reasonable agreement between the preliminary experimental results and the behaviour of the world data is observed. \\
\hspace*{\parindent}The work has been supported in part by the Russian Foundation for Basis Research grant $No.$ 10-02-00087a.


\begin{thebibliography}{99}
\bibitem{ref_1}
C.~E.~Garlson, J.~R.~Hiller, R.~J.~Holt, Ann. Rev. Nucl. Part. Sci. {\bfseries 47}, 395 (1997).

\bibitem{ref_2}
M.~Gar\c{c}on, J.~W.~Van Orden, Adv. Nucl. Phys. {\bfseries 26}, 293 (2001).

\bibitem{ref_3}
R.~Gilman, F.~Gross, nucl-th/0111015 (2001).

\bibitem{ref_7_src_ladygin}  W.~Gl$\ddot o$ckle, H.~Witala, D.~H$\ddot u$ber, H.~Kamada, J.~Golak, Phys. Rep. {\bfseries 274}, 107 (1996).

\bibitem{ref_8_src_ladygin} N.~Sakamoto et al., Phys. Lett. {\bfseries B367}, 60 (1996).


\bibitem{ref_10_src_ladygin} K.~Sekiguchi et al., Phys. Rev. {\bfseries C70}, 014001 (2004).

\bibitem{ref_11_src_ladygin} K.~Hatanaka et al., Phys. Rev. {\bfseries C66} 044002 (2002). 


\bibitem{ref_14_src_ladygin} K.~Ermisch et al., Phys. Rev. {\bfseries C68}, 051001 (2003). 

\bibitem{ref_const_count_roles1} S.~J.~Brodsky, G.~R.~Farrar, Phys. Rev. Lett., {\bfseries 31}, 1153 (1973).

\bibitem{ref_const_count_roles2} V.~A.~Matveev, R.~M.~Muradyan and A.~N.~Tavkhelidze, Lett. Nuovo Cimento {\bfseries 7}, 719 (1973).

\bibitem{ref_uzikov} Yu.~N.~Uzikov, JETP Lett., {\bfseries 81}, 387 (2005)

\bibitem{ref_2_geydar} R.~Schicker et al., Nucl. Instr. Meth. {\bfseries A380}, 586 (1996).

\bibitem{ref_4_geydar} P.~Salabura et al., Prog. Part. Nucl. Phys. {\bfseries 53}, 49 (2004).

\bibitem{ref_5_geydar} J.~Friese et al., Prog. Part. Nucl. Phys. {\bfseries 42}, 235 (1999).

\bibitem{ref_hades_detector_overview} G.~Agakishiev et al., Eur. Phys. J. {\bfseries A41}, 243 (2009).

\bibitem{ref_1_data} E.~Winkelman et al., Phys. Rev. C, {\bfseries 21}, 2535 (1980). 
\bibitem{ref_2_data} G.~W.~Bennet et al., Phys. Rev. Lett. {\bfseries 19}, 387 (1967).
\bibitem{ref_3_data} E.~Coleman et al., Phys. Rev. Lett, {\bfseries 16}, 761 (1966).
\bibitem{ref_4_data} N.~E.~Both et al., Phys. Rev. D, {\bfseries 4}, 1261 (1971).
\bibitem{ref_5_data} E.~Gulmez et al., Phys. Rev. C, {\bfseries 5}, 2067 (1991). 
                        


\end{thebibliography}
\end{document}